\documentclass[aps,pra,twocolumn,superscriptaddress]{revtex4}

\usepackage{amsmath,amsfonts,amssymb}
\usepackage{graphicx,float,calc}
\usepackage{color,bm}
\usepackage{ulem}
\usepackage{braket}
\usepackage{hyperref}
\usepackage{graphicx,epstopdf}

\newcommand{\ehbar}{\hbar_{\text{eff}}}

\newcommand{\mtn}{\mathcal{N}}

\begin{document}
%\begin{CJK}{GBK}{song}  %%%% begin Chinese, Japanese, and Korea languages environment

\title{Chaotic dynamics of complex trajectory and its quantum signature}

\author{Wen-Lei Zhao}
\email[]{wlzhao@jxust.edu.cn}
\affiliation{
School of Science, Jiangxi University of Science and Technology, Ganzhou 341000, China
}
\author{Pengkai Gong}
\affiliation{
School of Science, Jiangxi University of Science and Technology, Ganzhou 341000, China
}
\author{Jiaozi Wang}
\affiliation{
Department of Modern Physics, University of Science and Technology of China, Hefei 230026, China
}
\author{Qian Wang}
\email[]{qwang@zjnu.edu.cn}
\affiliation{
Department of Physics, Zhejiang Normal University, Jinhua 321004, China
}

\begin{abstract}
We investigate both the quantum and classical dynamics of a non-Hermitian system via a kicked rotor model with $\mathcal{PT}$ symmetry. For the quantum dynamics, both the mean momentum and mean square of momentum exhibits the staircase growth with time when the system parameter is in the neighborhood of the $\mathcal{PT}$ symmetry breaking point.
If the system parameter is very larger than the $\mathcal{PT}$ symmetry breaking point, the accelerator mode results in the directed spreading of the wavepackets as well as the ballistic diffusion in momentum space. For the classical dynamics, the non-Hermitian kicking potential leads to exponentially-fast increase of classical complex trajectories. As a consequence, the imaginary part of trajectories exponentially diffuses with time, while the  real part exhibits the normal diffusion. Our analytical prediction of the exponential diffusion of imaginary momentum and its breakdown time is in good agreement with numerical results. The quantum signature of the chaotic diffusion of the complex trajectories is reflected by the  dynamics of the out-of-time-order correlators (OTOC). In the semiclassical regime, the rate the exponential increase of the OTOC is equal with that of the exponential diffusion of complex trajectories.

\vspace{0.5cm}{\hspace{-0.3cm}{\bf Keywords:} $\cal{PT}$ symmetry, quantum-classical correspondence, quantum chaos}

\end{abstract}
\date{\today}

\pacs{03.65.-w, 03.65. YZ, 05.45.-a, 05.45.Mt}

%
% 05.45.-a Nonlinear dynamics and chaos
%
% 05.60.Gg Quantum transport
% 03.75.Kk Dynamic properties of condensates; collective and hydrodynamic excitations, superfluid flow
% 37.10.Jk Atoms in optical lattices
%

\maketitle

%%%%%%%%%%%%%%%%%%%%%%%%%%%%%%%%%%%%%%%%%%%%%%%%%%%%%%%%%%%%

\section{Introduction}
Traditional quantum mechanics requires that every physical observable be represented by a Hermitian operator, so as to ensure the observable a completely real eigenspectrum. However, the seminal work of Bender and Boettcher~\cite{Bender1998}  proved that Hemiticity is actually a sufficient but not necessary condition to guarantee the real spectrum. Since then, non-Hermitian quantum mechanics has inspired a great deal of work in various fields of quantum physics~\cite{Bender2002,Mosta2002,Bender2007,Mosta2010,Harsh2010,Moiseyev2011,Harsh2014,Cao2015,Ganainy2018,Berry2004,Klaiman2008,Graefe2008,Mosta2009,Graefe2011,Hou2017,Joshi2018}. Non-Hermitian quantum mechanics has also been used to describe the reduced open quantum systems~\cite{Rotter2009,Sergi2014}.
In particular, non-Hermitian features have been visualized in many experiments~\cite{Choi2010,Ott2013,Ruter2010,Longhi2010}. We can obtain a special type of non-Hermitian systems from those with  $\mathcal{PT}$-symmetric Hamiltonian~\cite{Buslaev1993,Bender2005}, whose striking feature is that they possess an entirely real-valued spectrum below the symmetry-breaking point, albeit non-Hermitian. Moreover, various optical structures have been proposed to test and realize the unique properties of $\mathcal{PT}$-symmetric Hamiltonian~\cite{Muga2005,Ganainy2007,Longhi2009,Guo2009,Feng2011,Alexeeva2012,
Regensburger2012}.

On the other hand, the research of periodically-driven systems has become increasingly popular in recent years.  Many intrinsic quantum phenomena, such as many-body localization~\cite{Huse2015,Ponte2015,Lazarides2015}, time crystals~\cite{Zhang2017,Yao2017,Yao2018,Nayak2018} and Floquet topological phase~\cite{Lindner2011,Titum2015,Zhou2016,Roy2017}, have been revealed in this type of systems. Nowadays, periodically-driven systems have a wide range of applications and play an important role in many branches of physics~\cite{Grifoni1998,Della2007,Della2013,Huang2020a,Huang2020b}. Therefore, it is of great interest to extend the research of periodically-driven quantum systems to non-Hermitian quantum mechanics. So far, various population oscillations~\cite{Wu2012,Ganainy2012}, abnormal level crossing rule~\cite{Moiseyev22011} and exotic topological phases~\cite{Gong2013,Zhou2018} have been found in the time-periodic non-Hermitian systems with $\mathcal{PT}$ symmetry. However, more work is needed to get a better insight into periodically-driven $\mathcal{PT}$-symmetric systems.

In this context, we investigate the chaotic dynamics of a non-Hermitian kicked rotor model with the kicking potential being $\mathcal{PT}$ symmetry~\cite{West2010,Longhi2017,Zhao19}.
As a paradigm model in the studies of periodically-driven systems, the quantum kicked rotor has been investigated in several works~\cite{Haake2010,Casati1979,Casati1987,Izrailev1990,Wang2011,Liujie06,Fishman1982,Podolskiy2004,Raizen1999,Rosen2000,Chaudhury2009,Zhao20,Zhao2010,Zhao09,Yang15}.
We will reveal the chaotic features of
the $\mathcal{PT}$-symmetric kicked rotor ($\mathcal{PT}$KR) model via its dynamical properties.
We find that depending on the strength of the imaginary part of the kicking potential, the $\mathcal{PT}$KR shows different dynamics. Namely, if the strength of the imaginary part  is in the neighborhood of $\mathcal{PT}$-symmetry breaking point, the transition between quasieigenstates leads to the jump of both the mean momentum and mean square of momentum. For very strong imaginary kicking potential, the non-Hermitian kicking potential produces stable wavepacket which spreads unidirectionally and diffuses ballistically with time.

The classical dynamics of non-Hermitian systems is assumed to be governed by the Hamiltonian principle for which the trajectory is complex~\cite{Graefe15,Bender2007,Bender2009,Bender09bk}.
It is previously found that the dynamical behavior of complex trajectories captures the feature of $\mathcal{PT}$-symmetry phase transition~\cite{Bender99} and dominates the quantum tunnelling~\cite{Bender11}. Therefore, the dynamics of complex trajectories is of significantly importance in many fields of physics. In the present work, we define the second moment (SM) for the real and imaginary parts of momentum, respectively, in order to quantify the classical diffusion. In chaotic situation, the SM of the real momentum increases linearly with time, while that of the imaginary momentum increases exponentially. Interestingly, they all exhibit a sharp transition to saturation level at a threshold time $\tau$. The underlying physics is due to the exponentially-fast increase of complex trajectories. Our theoretical prediction of the exponential increase of the SM of imaginary momentum and the critical time $\tau$ is in good agreement with numerical results.

To further demonstrate the chaotic features in the quantum dynamics, we assess the
out-of-time-order correlators (OTOC) \cite{Larkin1969,Maldacena2016}, which as a measurement of the dynamical instability in quantum chaos has been widely explored, both theoretically
(see, e.g., Refs.~\cite{Hashimoto2017,Dora2017,Heyl2018,Mata2018,Mata2018B,Fortes2019,Ueda2018,Yan19} and references therein) and experimentally \cite{Swingle2016,Hafezi2016,Li2017,Garttner2017}. We find that in the semi-classical regime, the dynamics of OTOC echoes its classical counterpart within the Ehrenfest time interval, both of which grows exponentially with time. The quantum-classical correspondence of OTOC proves the viability of the extension of the Hamiltonian principle to non-Hermitian systems. More important is that the rate of the exponential growth of OTOC equals to that of the SM of imaginary trajectories. Therefore, it is convincing that quantum OTOC is a signature of the classically-chaotic diffusion of complex trajectories.

The article is organized as follows. 
In Sec.~\ref{QDiffSect}, we describe our model and show the quantum diffusion. The classically-chaotic diffusion is presented in Sec.~\ref{CSect}. In Sec.~\ref{QSect}, we reveal the quantum signature of the chaotic diffusion  via the dynamics of the OTOC. Summary is presented in Sec.~\ref{Sum}.

\section{Quantum diffusion in momentum space}\label{QDiffSect}
We consider the $\mathcal{PT}$KR model for which the Hamiltonian in
dimensionless units takes the form~\cite{Longhi2017,Zhao19}
\begin{equation}\label{Hamil}
{\rm H} = \frac{p^2}{2} + V(\theta)\sum_j \delta(t-j)\;,
\end{equation}
with
\begin{equation}\label{Kicking}
V(\theta) = K\left[ \cos(\theta) + i \lambda \sin(\theta)\right]\;,
\end{equation}
where $\theta$ is the angle coordinate,  $p$ is the angular momentum operator written as $p = -i\ehbar {\partial}/{\partial \theta}$, and $\ehbar$ is the effective Planck constant. On the basis of the angular momentum operator $p |\varphi_n \rangle = n \ehbar |\varphi_n \rangle$ with $\langle \theta |\varphi_n \rangle= e^{i n \theta}/\sqrt{2\pi}$, an arbitrary quantum state is expanded as
$|\psi \rangle = \sum_n \psi_n |\varphi_n \rangle$.
The time evolution of a quantum state is governed by the Floquet operator
\begin{equation}\label{evol}
U = \exp\left(-i \frac{p^2}{2\ehbar}\right) \exp\left[-i\frac{V(\theta)}{\ehbar}\right]\;.
\end{equation}
The eigenequation of the Floquet operator reads
\begin{equation}
U|\psi_{\varepsilon}\rangle =  e^{-i \varepsilon}|\psi_{\varepsilon}\rangle\;,
\end{equation}\label{EEQFloqoper}
where $\varepsilon$ is the quasienergy. An intrinsic property of this system is that the real quasienergy eigenvalues
become complex, i.e., $\varepsilon =\varepsilon_r \pm \varepsilon_i$, when the strength of the imaginary part of the complex potential exceeds a threshold value $\lambda_{c}$~\cite{Longhi2017,West2010,Zhao19}.
Such a phenomenon is named as the spontaneous $\mathcal{PT}$-symmetry breaking.
\begin{figure}
\begin{center}
\includegraphics[width=8.5cm]{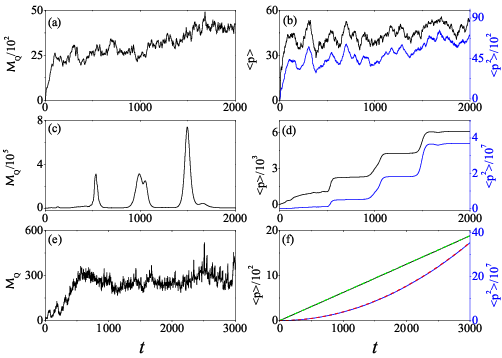}
\caption{(Color online) Left panels: Time dependence of the $\mathcal{M}_Q$. Right panels:
Time dependence of $\langle p \rangle$ (black lines) and $\langle p^2 \rangle$ (blue lines). In (f): Dash-dotted (in green) and dashed (in red) line denote the function of the form $\langle p \rangle \approx 2\pi t$ and $\langle p^2 \rangle \approx 4\pi^2 t^2$, respectively.
Parameters: $\lambda=0.02$ (top), 0.07 (middle) and 0.5 (bottom) with $K=7$ and $\ehbar=1.4$.
}\label{MeanVals}
\end{center}
\end{figure}

\begin{figure}
\begin{center}
\includegraphics[width=8.5cm]{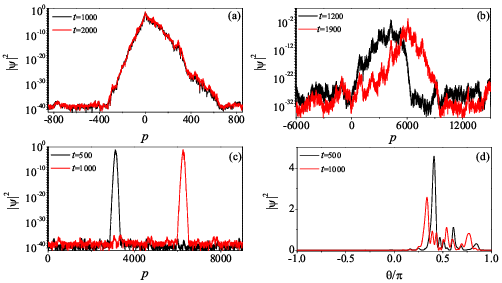}
\caption{(Color online) (a)-(c): Probability density distribution in angular momentum space with $\lambda=0.02$ (a), 0.07 (b) and 0.5 (c), respectively. (d): Probability density distribution in angle coordinate space for $\lambda=0.5$. The parameters are $K=7$ and $\ehbar=1.4$.
}\label{momentdistr}
\end{center}
\end{figure}

Quantum diffusion in momentum space is quantified by the SM of the wavepacket
\begin{equation}\label{smoment}
\mathcal{M}_Q(t)=\langle p^2 (t)\rangle -\left(\langle p(t)\rangle\right)^2\;,
\end{equation}
where $\langle p(t) \rangle = {\sum_n p_n |\psi_n(t)|^2}/{\mtn}
$ is the mean momentum, $\langle p^2(t) \rangle = {\sum_n p_n^2 |\psi_n(t)|^2}/{\mtn}$ is the mean square of the momentum, and $\mtn (t)=\sum_n |\psi_n(t)|^2$ is the norm of a quantum state. Note that this kind of definition of $\mathcal{M}_Q$ reduces the contribution from the norm which exponentially increases with time in the broken phase of $\mathcal{PT}$ symmetry.
We numerically investigate the $\mathcal{M}_Q(t)$ for a wide regime of $\lambda$, so that we can observe the rich physics resulting from the breaking of $\mathcal{PT}$ symmetry. In numerical simulations, the initial state is the ground state of the unperturbed Hamiltonian ${\rm H}_0=p^2/2$, i.e., $\psi_0(\theta)=1/\sqrt{2\pi}$.
Our numerical results show that, for small $\lambda$ [e.g., $\lambda=0.02$ in Fig.~\ref{MeanVals}(a)],
the $\mathcal{M}_Q$ increases during a very short time interval, after which it asymptotically unchanged with time evolution. Correspondingly, both the $\langle p\rangle $ and  $\langle p^2 \rangle$ saturate as time evolves [see Fig.~\ref{MeanVals}(b)], which demonstrates the appearance of dynamical localization.
In this situation, the wave function is exponentially localized in momentum space and quasiperiodically appears with time evolution [see Fig.~\ref{momentdistr}(a)]. More important is that such exponentially localized shape of wavepacet is asymmetric, for which the probability distribution in positive momentum is much larger. This leads to the positive values of mean momentum.  It is reasonable to believe that the mechanism of dynamical localization governs the quasiperiodic evolution of quantum states, when the $\mathcal{PT}$ symmetry is preserved for $\lambda< \lambda_c$.

We further investigate the wavepackets dynamics for the case that the value of $\lambda$ is in the vicinity of the phase-transition point, i.e., $\lambda \approx \lambda_c$. Our numerical results
show that the $\mathcal{M}_Q$ saturates rapidly with time evolution,  while there are some peaks occurring irregularly [see Fig.~\ref{MeanVals}(c)].
The corresponding mean values of both the $\langle p \rangle$ and  $\langle p^2 \rangle$ exhibit the stepwise growth with time [see Fig.~\ref{MeanVals}(d)]. Detailed observations find that each peak of $\mathcal{M}_Q$ corresponds to the jump of the $\langle p \rangle$ and  $\langle p^2 \rangle$ from lower stair to the upper one [see Fig.~\ref{MeanVals}(d)]. And the plateau of these two mean values corresponds to the saturation region of the $\mathcal{M}_Q$.
The underlying physics of such intrinsic phenomenon of the quantum diffusion can be revealed by the time evolution of wavepackets in momentum space.
The comparison of the quantum states at different time demonstrates that the momentum distribution has almost fixed width corresponding to the saturation of $\mathcal{M}_Q$ [see Fig.~\ref{momentdistr}(b)]. The spreading of the wavepackets to the positive direction in momentum space results in the increase of both the $\langle p \rangle$ and  $\langle p^2 \rangle$.
In fact, we previously found that for $\lambda\approx \lambda_c$, the quantum state corresponding to the appearance of the plateau of mean values is virtually one of the quasieignstate for which the imaginary part of the quasienergy $\varepsilon_i$ is significantly large. Moreover, the quasieignstates are exponentially-localized in momentum space. The transition of the quantum state between different quasieigenstates leads to the jumping of mean values~\cite{Zhao19}.

We also investigate the quantum diffusion for the case with $\lambda \gg \lambda_c$. Our results show that
$\mathcal{M}_Q(t)$ saturates very rapidly as time evolves [e.g., $\lambda=0.5$ in Fig.~\ref{MeanVals}(e)]. Correspondingly, the $\langle p \rangle$ increases linearly with time, i.e., $\langle p \rangle \propto t$, which leads to the ballistic diffusion of the energy, i.e., $\langle p^2 \rangle \propto t^2$ [see Fig.~\ref{MeanVals}(f)]. More interestingly, in the process of the linear acceleration of momentum,
the wavepacket has a stable shape and spreads to the positive direction in momentum space [see Fig.~\ref{momentdistr}(c)]. In angle coordinate space,
the wavepackets is mainly localized around the position of $\theta_0 = \pi/2$ [see Fig.~\ref{momentdistr}(d)]. This is due to the gain-or-loss mechanism of the non-Hermitian kicking potential, i.e., $U_{\lambda}=\exp[K\lambda \sin(\theta)/\ehbar]$~\cite{Zhao19}. The maximum value of $U_{\lambda}$ corresponds to $\theta_0$. After the action of $U_{\lambda}$, a quantum state is greatly enlarged within the neighborhood of $\theta_0$.
Remember that, the driven force of the real part of the kicking potential $K\cos(\theta)$ is positive for $\theta \in (0,\pi)$. Therefore, the quantum particle moves to positive direction.

\section{Chaotic diffusion of complex trajectory}\label{CSect}

We consider the $\mathcal{PT}$KR model for which the Hamiltonian in
dimensionless units takes the form~\cite{Longhi2017,Zhao19}
\begin{equation}\label{Hamil}
{\rm H} = \frac{p^2}{2} + V(\theta)\sum_j \delta(t-j)\;,
\end{equation}
with
\begin{equation}\label{Kicking}
V(\theta) = K\left[ \cos(\theta) + i \lambda \sin(\theta)\right]\;,
\end{equation}
where
$p$ is the angular momentum,  $\theta$ is the angle coordinate,  $K$ is the strength of the real part of the kicking potential, and  $\lambda$ indicates the strength of the imaginary part of the kicking potential. The kicking potential satisfies the condition of $\mathcal{PT}$ symmetry $V(\theta) = V^*(-\theta) $~\cite{West2010}. The extension of the Hamiltonian principle to this $\cal{PT}$KR model yields the classical mapping equation~\cite{Chirikov1979}
\begin{equation}\label{ClMap-MT}
\begin{cases}
p_{n+1} = p_n +K \left[ \sin(\theta_n) - i \lambda \cos(\theta_n)\right]\\
\theta_{n+1} = \theta_{n} + p_{n+1}
\end{cases}\;,
\end{equation}
where $\theta_n$ and $p_n$ separately denote angle coordinate and angular momentum after the $n$th kick.
It is worth noting that the extension of Hamiltonian canonical equation to non-Hermitian system has been widely investigated in different fields of physics~\cite{Graefe15,Bender2007,Bender2009,Bender09bk,Bender99,Bender11}. It is previously reported that the complex trajectories, for which the dynamics is governed by the Hamiltonian canonical equation, provide a new solution to brachistochrone problem~\cite{Bender09bk}. The dynamical behavior of complex trajectories captures the feature of $\cal{PT}$-symmetry phase transition~\cite{Bender99}. Quantum tunnelling is an anomaly of classical tunnelling of complex trajectories~\cite{Bender11}.
Accordingly, the classical trajectory should be  complex
\begin{equation}\label{CTraj-MT}
\begin{cases}
p_{n} = p_{n}^{r} + i p_{n}^{i} \\
\theta_{n} = \theta_{n}^{r} + i \theta_{n}^{i} \\\end{cases}\;,
\end{equation}
where $p_{n}^{r}$ and $p_{n}^{i}$ denote the real and imaginary parts of momentum $p_{n}$, respectively, $\theta_{n}^{r}$ and $\theta_{n}^{i}$ are that of coordinate $\theta_{n}$.
By substituting Eq.~\eqref{CTraj-MT} into Eq.~\eqref{ClMap-MT}, we get the mapping equations for complex trajectory
%\begin{widetext}
\begin{equation}\label{Mapping}
\begin{split}
p_{n+1}^{r} &= p_{n}^{r} + K \sin(\theta_{n}^{r}) \left[\cosh(\theta_{n}^{i})-\lambda \sinh(\theta_{n}^{i})\right]\;,\\
p_{n+1}^{i} &= p_{n}^{i} +  K\cos(\theta_{n}^{r})\left[\sinh(\theta_{n}^{i})-\lambda \cosh(\theta_{n}^{i})\right]\;,\\
\theta_{n+1}^{r} & =  \theta_{n}^{r} +  p_{n+1}^{r}\;,\\
\theta_{n+1}^{i}& =  \theta_{n}^{i} + p_{n+1}^{i}\;.
\end{split}
\end{equation}
%\end{widetext}

\begin{figure}[htbp]
\begin{center}
\includegraphics[width=8.5cm]{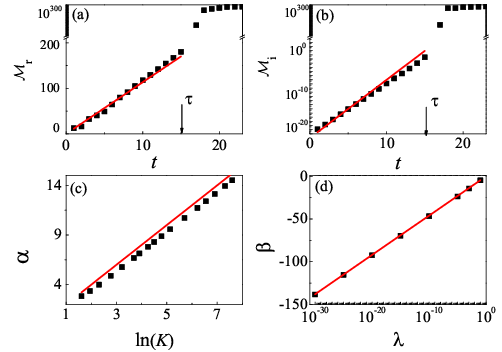}
\caption{(color online).  Top panels: Time dependence of the $\mathcal{M}_r$ (a) and $\mathcal{M}_i$ (b). In (a): Red line indicates the normal diffusion of the form $\mathcal{M}_r(t) = D t$ with $D\approx K^2 /2$. Arrow marks the transition points $\tau$ beyond which the time evolution of $\mathcal{M}_r$ departs from the normal diffusion. In (b): Red line indicates the analytic expression of the form $\mathcal{M}_i(t) \approx \exp (\alpha t + \beta)$ in Eq.~\eqref{Expdiffusion}.
Arrow indicates the transition point $\tau$ for which the time growth of $\mathcal{M}_i$ is faster than the exponential increase. The parameters are $K=5$ and $\lambda = 10^{-10}$.
(c) The growth rate $\alpha$ versus $\ln(K)$ for $\lambda = 10^{-10}$. Red line indicates the analytic prediction  $\alpha = 2\ln(K)$ in Eq.~\eqref{EXPRate}. (d) The value of $\beta$ versus $\ln(\lambda)$ with $K = 5$. Red line  indicates the analytic expression $\beta = 2\ln(\lambda)$ in Eq.~\eqref{EXPFactor}. \label{diffusion}}
\end{center}
\end{figure}

Based on the above equations, we investigate the classical dynamics of the $\mathcal{PT}$KR model.
In order to quantify the classical diffusion, we define the
SM (or variance) for the real and imaginary parts of  momentum as
\begin{equation}\label{SMTReal}
\mathcal{M}_r(t) = \langle p_r^2(t) \rangle - (\langle p_r(t) \rangle)^2\;,
\end{equation}
and
\begin{equation}\label{SMTImag}
\mathcal{M}_i(t) = \langle p_i^2(t) \rangle - (\langle p_i(t) \rangle)^2\;.
\end{equation}
In numerical simulations, we set the initial values of classical trajectories as $p_i = p_r =0$, $\theta_i =0$, and $\theta_r$ being random variables uniformly distributed in the interval $[-\pi,\pi]$. The total number of trajectories is $N=10^5$.
We find that the SM of the real part of momentum exhibits the normal diffusion $\mathcal{M}_r(t) = D t$ with $D \approx K^2/2$ for  time smaller than a threshold value, i.e.,  $t<\tau$. Beyond such a threshold time $t > \tau$, the $\mathcal{M}_r(t)$ exhibits a sharp transition to saturation (see Fig.~\ref{diffusion}(a)). Interestingly, the SM of the imaginary part of momentum  exponentially increases  $\mathcal{M}_i(t)\approx \exp(\alpha t + \beta)$ for  $t<\tau$, and it also saturates rapidly  if $t>\tau$ (see Fig.~\ref{diffusion}(b)).
We further numerically investigate the dependence of the growth rate $\alpha$ and the factor $\beta$ on system parameters. Our numerical result demonstrates the rule of the form $\alpha = 2\ln(K)$ and $\beta = 2\ln(\lambda)$ (see Fig.~\ref{diffusion}(c) and \ref{diffusion}(d)).

\begin{figure}[t]
\centering
\includegraphics[width=8.0cm]{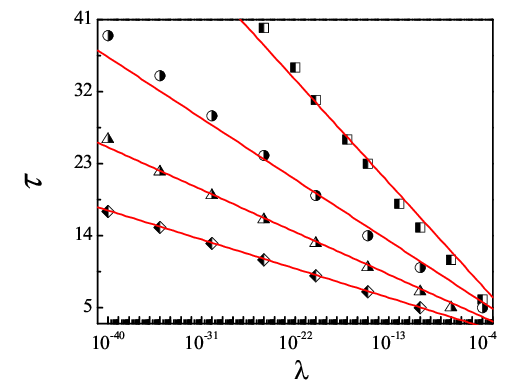}
\caption{(color online). Threshold time $\tau$ versus $\lambda$  with $K=5$ (squares), 15 (circles), 50 (triangles) and 300 (diamonds).
Red lines denote the analytic expression $\tau \approx - \log_K(\lambda)$ in Eq.~\eqref{CriticT2-MT}. \label{onetraj}}
\end{figure}

In fact, the classical diffusion is closely related to the exponential increase of complex trajectories. For example, we consider a special trajectory with initial value
$p_{0}^r = p_{0}^i=0$ and $\theta_{0}^r = \theta_{0}^i =0$.
It is easy to prove that, at an arbitrary time $t = n$, the real part of both the angle coordinate  and angular momentum of this trajectory is zero, i.e., $p_{n}^r =0$ and $\theta_{n}^r =0$, while its imaginary part exponentially increases, i.e., $p_{n}^i =\theta_{n}^i = - \lambda K^n$~\cite{analysis}. Therefore, it is reasonable to believe that the SM increases in the way
\begin{equation}\label{Expdiffusion}
\mathcal{M}_i(t)\approx (\lambda K^n)^2=\exp(\alpha t + \beta)\;,
\end{equation}
where the growth rate $\alpha$ only depends on $K$
\begin{equation}\label{EXPRate}
\alpha = 2\ln(K)\;,
\end{equation}
and the factor $\beta$ is the function of $\lambda$
\begin{equation}\label{EXPFactor}
\beta = 2\ln(\lambda)\;.
\end{equation}
Our analytical prediction is confirmed by numerical results, as shown in Figs.~\ref{diffusion}(c) and (d)). Note that in the
derivation of $p_{n}^{\rm i}$ and $\theta_{n}^{\rm i}$ we have used the  condition
\begin{equation}\label{conditionNT-MT}
K^{t-1}\lambda \ll 1\;.
\end{equation}
Such exponential increase breaks down if $K^{t}\lambda \sim 1$. Thus, a rough estimation for the threshold time is
\begin{equation}\label{CriticT2-MT}
\tau \approx - \log_K(\lambda)\;.
\end{equation}
In order to confirm the above analysis, we numerically investigate the threshold time $\tau$ for different $K$ and $\lambda$, which is depicted in Fig.~\ref{onetraj}. It is seen that our numerical results are in good agreement with the analytical prediction in Eq.~\eqref{CriticT2-MT}, which again demonstrates the exponential growth of classical trajectories dominates the chaotic diffusion of this non-Hermitian system.

\section{Quantum-classical correspondence in terms of OTOC}\label{QSect}

In recent years, both theoretical~\cite{Hashimoto2017,Dora2017,Heyl2018,Mata2018,Mata2018B,Fortes2019,Ueda2018,Yan19}
and experimental~\cite{Swingle2016,Hafezi2016,Li2017,Garttner2017} investigations show that the OTOC $C(t) = \langle [\hat{A}(t), \hat{B}]^2\rangle$ is an effective indicator of chaos in quantum systems.
We consider the case that both $\hat{A}$ and $\hat{B}$ are angular momentum operators, hence $C(t) =  \left\langle [{p}(t), {p}]^2 \right\rangle$~\cite{Ueda2018,Rozenbaum17}. Note that, for $\cal{PT}$-symmetric systems, the norm $\cal{N}$ of quantum states may exponentially increase with time. To reduce the contribution of the norm to OTOC, we define the OTOC as
\begin{equation}\label{QOTOC}
C(t) = \frac{1}{\cal{N}} \left\langle [{p}(t), {p}]^2 \right\rangle\;.
\end{equation}
In the semi-classical limit, the quantum OTOC of Hermitian systems is in consistence with its classical counterpart $C(t)= \ehbar^2 C_{cl}(t)$ for time shorter than the Ehrenfest time $t<t_E$~\cite{Rozenbaum17}. The classical OTOC is expressed in the Possion bracket
\begin{equation}\label{COTOC}
C_{cl}(t) =\left\langle\{p(t),p(0)\}^2\right \rangle_{cl}= \left\langle \left( \frac{\partial p(t)}{\partial x(0) }\right)^2\right \rangle_{cl}\;,
\end{equation}
where $\langle \cdots \rangle_{cl}$ denotes the ensemble average over classical trajectories.
In numerical simulations, the classical OTOC is approximated by $C_{cl}(t)\approx \left\langle  {\delta p^2(t)}/{\delta x^2(0) }\right \rangle_{cl}$~\cite{Rozenbaum17}, where $\delta p(t)$ is the difference of the momentum of two trajectories at time $t$,  $\delta x(0)$ is deviation of the angle coordinate at the initial time.

For the $\mathcal{PT}$KR model, the classical trajectories are complex~\cite{Bender2007,Bender2009,Bender09bk,Graefe15,Bender99,Bender11}, thus $\delta p = \delta p_r + i \delta p_i$ and  $\delta \theta = \delta \theta_r + i \delta \theta_i$. Accordingly, a reasonable extension of the classical OTOC is
~\cite{analysis}
\begin{align}\label{OTOC-semi2}
C_{cl}(t) = \left\langle  \frac{\left|\delta p(t)\right|^2}{\left|\delta \theta(0)\right|^2 }\right \rangle_{cl}\;,
\end{align}
where $|\cdots|$ denotes the modular square of complex variables. Based on the classical mapping equations in Eq.~\eqref{Mapping}, we get the tangent mapping equations for the deviation of two trajectories
\begin{widetext}
\begin{equation}
\begin{split}
\delta p_{n+1}^{\rm r}   =& \delta p_{n}^{\rm r} +K \sin(\theta_{n}^{\rm r})\left[ \sinh({\theta}_{n}^{\rm i}) - \lambda \cosh({\theta}_{n}^{\rm i}) \right]\delta{\theta}_{n}^{\rm i}
+ K \cos({\theta}_{n}^{\rm r})\left[\cosh({\theta}_{n}^{\rm i})-\lambda \sinh({\theta}_{n}^{\rm i})\right]\delta{\theta}_{n}^{\rm r}\;,\\
\delta p_{n+1}^{\rm i}   =& \delta p_{n}^{\rm i} + K\cos(\theta_{n}^{\rm r})\left[  \cosh({\theta}_{n}^{\rm i})-\lambda \sinh({\theta}_{n}^{\rm i})  \right]\delta{\theta}_{n}^{\rm i}
- K\sin({\theta}_{n}^{\rm r}) \left[\sinh({\theta}_{n}^{\rm i})-\lambda \cosh({\theta}_{n}^{\rm i})\right]\delta{\theta}_{n}^{\rm r}\;,\\
\delta\theta_{n+1}^{\rm r}  = & \delta\theta_{n}^{\rm r} +  \delta p_{n+1}^{\rm r}\;,\\
\delta\theta_{n+1}^{\rm i} =  &\delta\theta_{n}^{\rm i} + \delta p_{n+1}^{\rm i}\;,
\end{split}
\end{equation}
\end{widetext}
where the superscript `$r(i)$' denotes the real (imaginary) part of a complex variable, and the subscript `$n$' indicates the time $t=n$~\cite{analysis}.
In numerical simulations, we set the initial values as  $\delta p_0^r = \delta p_0^i=0$, $\delta \theta_0^r = |\delta \theta_0| \cos(\phi)$ and $\delta \theta_0^i = |\delta \theta_0| \sin(\phi)$ with $|\delta \theta_0|= 10^{-10}$ and $\phi = \pi/4$.
For quantum simulations, the initial state is set as a Gaussian function
$\psi(\theta) = (\frac{\sigma}{\pi})^{1/4} \exp(-{\sigma}\theta^2/{2})$ with $\sigma = 10$.

\begin{figure}[htbp]
\begin{center}
\includegraphics[width=8.0cm]{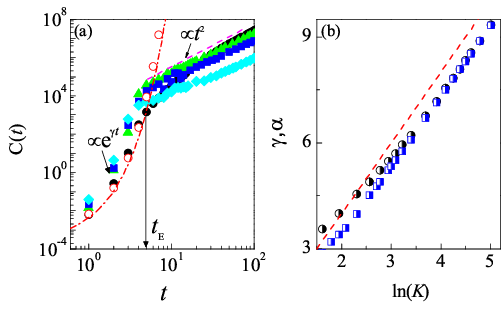}
\caption{(color online). (a) Time dependence of $C(t)$ (filled symbols) for $\lambda =10^{-5}$ (circles), $0.005$ (triangles), $0.007$ (squares) and $0.01$ (diamonds). The parameters are $K=8$ and $\ehbar=0.01$. Dash line denotes the power-law increase of the form  $C(t)\propto t^{2}$. In comparison, the empty circles denote the product of the classical OTOC with $\ehbar^2$, i.e., $\ehbar^2 C_{cl}(t)$ for $\lambda =10^{-5}$. Dash-dotted line (in red) denotes the exponential increases of the form $C(t)\propto e^{\gamma t}$ with $\gamma =2 \ln(K)$.
Arrow marks the Ehrenfest time $t_E$ for $\lambda = 10^{-5}$.
(b) Comparison of $\alpha$ (circles) and $\gamma$ (squares) for different $K$ with  $\lambda=10^{-10}$ and $\ehbar=0.001$. Red dashed line indicates our analytic prediction $\alpha = 2 \ln(K)$ in Eq.~\eqref{EXPRate}.\label{OTOCSemiC}}
\end{center}
\end{figure}

Our investigation shows that, for very small $\lambda$ (e.g., $\lambda = 10^{-5}$), the quantum OTOC is in good agreement with its classical counterpart during the Ehrenfest time $t< t_E$ [see Fig.~\ref{OTOCSemiC}(a)]. Both of them increase exponentially with time $C(t)\propto e^{\gamma t}$. Such quantum-classical correspondence demonstrates that the extension of the Hamiltonian equation to non-Hermitian systems is valid.
For $t > t_E$, the time evolution of quantum OTOC exhibits a clear transition to the power-law increase $C(t)\propto t^2$, which is similar to that of the Hermitian case~\cite{Ueda2018}. In addition, we numerically investigate the dynamics of OTOC for different $\lambda$, which is qualitatively the same [see Fig.~\ref{OTOCSemiC}(a)].
As a further step, we investigate the growth rate $\gamma$ of $C(t)$ for different $K$. Interestingly, the  growth rate $\gamma$ equals to that of $\mathcal{M}_i(t)$, i.e., $\gamma = \alpha$ [see Fig.~\ref{OTOCSemiC}(b)]. Therefore, we can believe that the quantum OTOC is a signature of the chaotic diffusion of complex trajectory.

\section{Summary}\label{Sum}

In this work, we make detailed investigations on both the classical and quantum dynamics of the $\mathcal{PT}$KR model. {For $\lambda \approx \lambda_c$, the SM of the
wavepacket in momentum space $\mathcal{M}_Q$ is asymptotically unchanged with time evolution, expect some peaks occurring irregularly. This is due to the fact that the quantum states evolve to the quasieigenstates which is exponentially localized at different position in  momentum space. The transition between different quasieigenstates leads to the irregular occurrence of the peaks of $\mathcal{M}_Q$. In this situation, both the $\langle p(t)\rangle $ and $\langle p^2(t)\rangle$ exhibit the staircase growth with time.
For $\lambda \gg \lambda_c$, the non-Hermitian kicking potential produces the stabilized wavepakets, which accelerates unboundedly towards positive direction, i.e., $\langle p(t)\rangle \propto t $. In this situation, the quantum particle diffuses ballistically i.e., $\langle p^2(t)\rangle \propto t^2 $ with fixed value of  $\mathcal{M}_Q$.}

For the classical dynamics, some trajectories exponentially increase
with time, which leads to the exponentially-fast diffusion of imaginary momentum, i.e.,
$\mathcal{M}_i(t)\approx \exp(\alpha t + \beta)$ with  $\alpha = 2\ln(K)$  and $\beta = 2\ln(\lambda)$. The real part of the momentum exhibits the normal diffusion, i.e., $\mathcal{M}_r = Dt$. The classical diffusion for both the real and imaginary momentum breaks down at a threshold time which depends on the system parameter in the form of $\tau \approx -\log_K(\lambda)$.
The quantum signature of chaotic diffusion of complex trajectories is reflected by the dynamics of the OTOC $C(t)$. In the semiclassical regime,  the OTOC increases exponentially with time  $C(t) \propto e^{\gamma t}$ for time smaller than the Ehrenfest time, i.e., $t < t_E$, after which it increases with the power law, i.e., $C(t) \propto t^2$. The growth rate $\gamma$  of $C(t)$ equals to that of $\mathcal{M}_i(t)$, which demonstrates that the OTOC is the fingerprint of chaotic diffusion of complex trajectories.
Our investigation is helpful for understanding the quantum-classical correspondence in non-Hermitian systems.

\section*{ACKNOWLEDGMENTS}
This work was partially supported by the Natural Science Foundation of China under Grant Nos. 12065009, 11804130 and 11805165. Zhejiang Provincial Nature Science Foundation under Grant No. LY20A050001.

\appendix

\section{Exponentially-fast growth of a special trajectory}

The classical mapping equations for the complex trajectory read
\begin{equation}\label{MappingAPX}
\begin{split}
p_{n+1}^{\rm r} &= p_{n}^{\rm r} + K \sin(\theta_{n}^{\rm r}) \left[\cosh(\theta_{n}^{\rm i})-\lambda \sinh(\theta_{n}^{\rm i})\right]\;,\\
p_{n+1}^{\rm i} &= p_{n}^{\rm i} +  K\cos(\theta_{n}^{\rm r})\left[\sinh(\theta_{n}^{\rm i})-\lambda \cosh(\theta_{n}^{\rm i})\right]\;,\\
\theta_{n+1}^{\rm r} & =  \theta_{n}^{\rm r} +  p_{n+1}^{\rm r}\;, \\
\theta_{n+1}^{\rm i}& =  \theta_{n}^{\rm i} + p_{n+1}^{\rm i}\;.
\end{split}
\end{equation}
Let us consider  a special trajectory with the initial values $(\theta_{0}^{\rm r}=0, \theta_{0}^{\rm i}=0, p_{0}^{\rm r}=0, p_{0}^{\rm i}=0)$. The above mapping equation yields the trajectory, at the time $t=1$,  $p_{1}^{\rm r} =\theta_{1}^{\rm r} = 0$, $p_{1}^{\rm i} =\theta_{1}^{\rm i}= -K\lambda$. By repeating the same derivation, one can find that
the real part of the trajectory  at the time $t=2$ is zero,
$p_{2}^{\rm r} =\theta_{2}^{\rm r}= 0$. The imaginary part of momentum has the expression $p_{2}^{\rm i} = -K\lambda + K\left[\sinh(-K\lambda)-\lambda \cosh(-K\lambda)\right]$.
In condition that $K \lambda \ll 1$ and $K\gg 1$, the $p_{2}^{\rm i}$ is approximated as
\begin{align}\label{PI2T1}
p_{2}^{\rm i}
&\approx -2K\lambda - K^2\lambda \approx  - K^2\lambda\;,
\end{align}
where we use the approximation $\sinh(K\lambda) \approx K\lambda$ and $\cosh(K\lambda) \approx 1$ for $K \lambda \ll 1$, and neglect the first term $-2K \lambda$ on the right of the above equation, since $K\lambda \ll K^2\lambda$ for $K \gg 1$.
The imaginary part of the angle coordinate reads
\begin{align}
\theta_{2}^{\rm i}& =  \theta_{1}^{\rm i} + p_{2}^{\rm i} = -K\lambda - K^2\lambda \approx - K^2\lambda\;.\label{ThetI2T}
\end{align}

It is straightforward to get the real parts of classical trajectory at $t=3$, i.e., $p_{3}^{\rm r} = 0$ and $\theta_{3}^{\rm r}  =  0$.
The imaginary part of the momentum is
\begin{align*}
p_{3}^{\rm i} &= -K^2\lambda + K\left[\sinh(-K^2\lambda)-\lambda \cosh(-K^2\lambda)\right]\;.
\end{align*}
In condition that $K^2\lambda \ll 1$ and $K\gg1$, the $p_{3}^{\rm i}$ is estimated as
\begin{align}\label{PI3T1}
p_{3}^{\rm i}
&\approx -K^2\lambda + K\left(-K^2\lambda-\lambda \right)\approx -K^3\lambda\;.
\end{align}
The imaginary part of the angle coordinate is expressed as
\begin{align}
\theta_{3}^{\rm i}& =  \theta_{2}^{\rm i} + p_{3}^{\rm i} = - K^2\lambda - K^3 \lambda \approx - K^3\lambda\;.\label{ThetI3T}
\end{align}

By repeating the same procedure
we get the trajectory at any time $t=n$, $p_{n}^{\rm r} =
\theta_{n}^{\rm r}=0$,
\begin{equation}\label{trajNT-APX}
\begin{split}
p_{n}^{\rm r}& =
\theta_{n}^{\rm r}=0\;,\\
p_{n}^{\rm i}&  \approx - K^n\lambda\;,\\
\theta_{n}^{\rm i}&  \approx - K^n\lambda\;.
\end{split}
\end{equation}
It is evident that,  the value of both $\theta_{t}^{\rm i}$ and  $p_{t}^{\rm i}$
exponentially grows with time. Note that in the derivation of $p_{n}^{\rm i}$ and $\theta_{n}^{\rm i}$ we have used the  condition
$K^{n-1}\lambda \ll 1$.
Such exponentially-fast growth will break once $K^{t}\lambda \approx 1$.

\section{Classical tangent mapping equation for calculating OTOC}
The classical dynamics is governed by the mapping equations in Eq.~\eqref{MappingAPX}.
Let us consider two complex trajectories
$(\theta_{n}^{\rm r},\theta_{n}^{\rm i},p_{n}^{\rm r},p_{n}^{\rm i})$ and $(\tilde{\theta}_{n}^{\rm r},\tilde{\theta}_{n}^{\rm i},\tilde{p}_{n}^{\rm i},\tilde{p}_{n}^{\rm i})$ for which the deviation is defined as
$\delta p_n^{\rm r}   = \tilde{p}_{n}^{\rm r}- p_{n}^{\rm r}$, $\delta p_n^{\rm i}   =  \tilde{p}_{n}^{\rm i} - p_{n}^{\rm i}$, $\delta \theta_n^{\rm r}  =   \tilde{\theta}_{n}^{\rm r} -\theta_{n}^{\rm r}$, and $\delta \theta_n^{\rm i}  =  \tilde{\theta}_{n}^{\rm i} - \theta_{n}^{\rm i}$.
From the classical mapping equations in Eqs.~\eqref{MappingAPX}, we get the mapping equations of the deviation
\begin{equation}\label{delt}
\begin{split}
\delta p_{n+1}^{\rm r}   =& \delta p_{n}^{\rm r} +K \sin(\tilde{\theta}_{n}^{\rm r}) \left[\cosh(\tilde{\theta}_{n}^{\rm i})-\lambda \sinh(\tilde{\theta}_{n}^{\rm i})\right]\\
&-K \sin(\theta_{n}^{\rm r}) \left[\cosh(\theta_{n}^{\rm i})-\lambda \sinh(\theta_{n}^{\rm i})\right]\;,\\
\delta p_{n+1}^{\rm i}   = &\delta p_{n}^{\rm i} + K\cos(\tilde{\theta}_{n}^{\rm r})\left[\sinh(\tilde{\theta}_{n}^{\rm i})-\lambda \cosh(\tilde{\theta}_{n}^{\rm i})\right]\\
&- K\cos(\theta_{n}^{\rm r})\left[\sinh(\theta_{n}^{\rm i})-\lambda \cosh(\theta_{n}^{\rm i})\right]\;,\\
\delta\theta_{n+1}^{\rm r}  =&  \delta\theta_{n}^{\rm r} +  \delta p_{n+1}^{\rm r}\;,\\
\delta\theta_{n+1}^{\rm i} =&  \delta\theta_{n}^{\rm i} + \delta p_{n+1}^{\rm i}\;.
\end{split}
\end{equation}
The evolution equations for $\delta p_{n}^{\rm r}$ and $\delta p_{n}^{\rm r}$
are nonlinear due to the sinusoidal (or cosinoidal) dependence on the $\tilde{\theta}_{n}^{\rm r}$.
For small $\delta{\theta}_{n}^{\rm r}$ and $\delta{\theta}_{n}^{\rm i}$, there are approximations that
$\sin(\tilde{\theta}_{n}^{\rm r})\approx  \sin({\theta}_{n}^{\rm r}) + \cos({\theta}_{n}^{\rm r})  \delta{\theta}_{n}^{\rm r}$, $\cos(\tilde{\theta}_{n}^{\rm r})\approx  \cos({\theta}_{n}^{\rm r}) - \sin({\theta}_{n}^{\rm r})  \delta{\theta}_{n}^{\rm r}$, $\sinh(\tilde{\theta}_{n}^{\rm i}) \approx \sinh({\theta}_{n}^{\rm i}) + \delta{\theta}_{n}^{\rm i}  \cosh({\theta}_{n}^{\rm i})$, and $\cosh(\tilde{\theta}_{n}^{\rm i}) \approx \cosh({\theta}_{n}^{\rm i}) + \delta{\theta}_{n}^{\rm i}  \sinh({\theta}_{n}^{\rm i})$. By straightforward calculation,  we get the mapping equations for $\delta p_{n}^{\rm r}$
and $\delta p_{n}^{\rm i}$
\begin{align*}
\delta p_{n+1}^{\rm r}   =& \delta p_{n}^{\rm r} +K \sin(\theta_{n}^{\rm r})\left[ \sinh({\theta}_{n}^{\rm i}) - \lambda \cosh({\theta}_{n}^{\rm i}) \right]\delta{\theta}_{n}^{\rm i}\\
 &+ K \cos({\theta}_{n}^{\rm r})\left[\cosh({\theta}_{n}^{\rm i})-\lambda \sinh({\theta}_{n}^{\rm i})\right]\delta{\theta}_{n}^{\rm r}\;,
\end{align*}
and
\begin{align*}
\delta p_{n+1}^{\rm i}   =& \delta p_{n}^{\rm i} + K\cos(\theta_{n}^{\rm r})\left[  \cosh({\theta}_{n}^{\rm i})-\lambda \sinh({\theta}_{n}^{\rm i})  \right]\delta{\theta}_{n}^{\rm i}\\
&- K\sin({\theta}_{n}^{\rm r}) \left[\sinh({\theta}_{n}^{\rm i})-\lambda \cosh({\theta}_{n}^{\rm i})\right]\delta{\theta}_{n}^{\rm r}\;.
\end{align*}
As a conclusion, the tangent mapping equations for the deviation of classical trajectories read
\begin{widetext}
\begin{equation}\label{TangentEQ}
\begin{split}
\delta p_{n+1}^{\rm r}   =& \delta p_{n}^{\rm r} +K \sin(\theta_{n}^{\rm r})\left[ \sinh({\theta}_{n}^{\rm i}) - \lambda \cosh({\theta}_{n}^{\rm i}) \right]\delta{\theta}_{n}^{\rm i}+ K \cos({\theta}_{n}^{\rm r})\left[\cosh({\theta}_{n}^{\rm i})-\lambda \sinh({\theta}_{n}^{\rm i})\right]\delta{\theta}_{n}^{\rm r}\;,\\
\delta p_{n+1}^{\rm i}   =& \delta p_{n}^{\rm i} + K\cos(\theta_{n}^{\rm r})\left[  \cosh({\theta}_{n}^{\rm i})-\lambda \sinh({\theta}_{n}^{\rm i})  \right]\delta{\theta}_{n}^{\rm i}- K\sin({\theta}_{n}^{\rm r}) \left[\sinh({\theta}_{n}^{\rm i})-\lambda \cosh({\theta}_{n}^{\rm i})\right]\delta{\theta}_{n}^{\rm r}\;,\\
\delta\theta_{n+1}^{\rm r}  = & \delta\theta_{n}^{\rm r} +  \delta p_{n+1}^{\rm r}\;,\\
\delta\theta_{n+1}^{\rm i} =  &\delta\theta_{n}^{\rm i} + \delta p_{n+1}^{\rm i}\;.
\end{split}
\end{equation}
\end{widetext}

In Hermitian systems, the classical OTOC is defined as~\cite{Rozenbaum17}
\begin{align}\label{OTOC-semi}
C_{cl}(t) = \left\langle \left[ \frac{\partial p(t)}{\partial \theta(0) }\right]^2\right \rangle_{cl}
\approx  \left\langle \left[ \frac{\delta p(t)}{\delta \theta(0) }\right]^2\right \rangle_{cl}\;.
\end{align}
We make an extension to non-Hermitian systems. Considering $\delta p(t) = \delta p_r(t) + i \delta p_i(t)$ and $\delta \theta(0) = \delta \theta_r(0) + i \delta \theta_i(0)$, it is reasonable to replace the square in the above equation with the modular square, i.e., $|\delta p(t)|^2 = |\delta p_r(t)|^2 +  |\delta p_i(t)|^2$ and $|\delta \theta(0)|^2 = |\delta \theta_r(0)|^2 +  |\delta \theta_i(0)|^2$.
Then, the OTOC has the expression
\begin{align}\label{OTOC-semi2}
C_{cl}(t) \approx \left\langle  \frac{\left|\delta p(t)\right|^2}{\left|\delta \theta(0) \right|^2}\right \rangle_{cl}\;.
\end{align}
It is known that, in the semi-classical limit, the quantum OTOC of Hermitian systems is consistent with its classical counterpart $C(t) \approx \ehbar^2 C_{cl}(t)$. We numerically simulate the time evolution of $(\delta \theta_n^{\rm r},\delta \theta_n^{\rm i},\delta p_n^{\rm r},\delta p_n^{\rm i})$ according to Eq.~\eqref{TangentEQ}, and thus   investigate the classical OTOC. Interestingly, the quantum-classical correspondence of OTOC exists in non-Hermitian systems, which proves that the extension  of  the Hamiltonian equation to non-Hermitian system is valid.

%\end{CJK*}  %% end Chinese, Japanese, and Korea languages environment


\begin{thebibliography}{*}
\bibitem{Bender1998}
Bender C M and Boettcher S 1998 \textit{Phys. Rev. Lett.} {\bf 80} 5243


\bibitem{Bender2002}
Bender C M, Brody D C and Jones H F 2002 \textit{Phys. Rev. Lett.} {\bf 89} 270401


\bibitem{Mosta2002}
Mostafazadeh A 2002 \textit{J. Math. Phys. (N. Y.)} {\bf 43} 2814

\bibitem{Bender2007}
Bender C M 2007 \textit{Rep. Prog. Phys.} {\bf 70} 947

\bibitem{Mosta2010}
Mostafazadeh A 2010 \textit{Int. J. Geom. Meth. Mod. Phys.} {\bf 07} 1191

\bibitem{Harsh2010}
Jones-Smith K and Mathur H 2010 \textit{Phys. Rev. A} {\bf 82} 042101


\bibitem{Moiseyev2011}
Moiseyev N 2011 {\it Non-Hermitian quantum mechanics} (Cambridge University Press, Cambridge, UK) p.211

\bibitem{Harsh2014}
Jones-Smith K and Mathur H 2014 \textit{Phys. Rev. D} {\bf 89} 125014


\bibitem{Cao2015}
Cao H and Wiersig J 2015 \textit{Rev. Mod. Phys.} {\bf 87} 61

\bibitem{Ganainy2018}
El-Ganainy R, Makris K G, Khajavikhan M, Musslimani Z H, Rotter S and Christodoulides D N 2018 \textit{Nat. Phys.} {\bf 14} 11


\bibitem{Berry2004}
Berry M 2004 \textit{Czech. J. Phys.} {\bf 54} 1039

\bibitem{Klaiman2008}
Klaiman S, G\"unther U and Moiseyev N 2008 \textit{Phys. Rev. Lett.} {\bf 101} 080402

\bibitem{Graefe2008}
Graefe E M, Korsch H J and Niederle A E 2008 \textit{Phys. Rev. Lett.} {\bf 101} 150408

\bibitem{Mosta2009}
Mostafazadeh A 2009 \textit{Phys. Rev. Lett.} {\bf  102} 220402

\bibitem{Graefe2011}
Graefe E -M and Schubert R 2011 \textit{Phys. Rev. A} {\bf 83} 060101(R)

\bibitem{Hou2017}
Hou T -J 2017 \textit{Phys. Rev. A} {\bf 95} 013824

\bibitem{Joshi2018}
Joshi S and Galbraith I 2018 \textit{Phys. Rev. A} {\bf 98} 042117

\bibitem{Rotter2009}
Rotter I 2009 \textit{J. Phys. A} {\bf 42} 153001

\bibitem{Sergi2014}
Zloshchastiev K G and Sergi A 2014 \textit{J. Mod. Opt.} {\bf 61} 1298

\bibitem{Choi2010}
Choi Y, Kang S, Lim S, Kim W, Kim J -R, Lee J -H and An K 2010 \textit{Phys. Rev. Lett.} {\bf 104} 153601

\bibitem{Ott2013}
Barontini G, Labouvie R, Stubenrauch F, Vogler A, Guarrera V and Ott H 2013 \textit{Phys. Rev. Lett.} {\bf  110} 035302


\bibitem{Ruter2010}
R\"uter C E, Makris K G, El-Ganainy R, Christodoulides D N, Segev M and Kip D 2010 \textit{Nat. Phys.} {\bf 6} 192

\bibitem{Longhi2010}
Longhi S 2010 \textit{Phys. Rev. Lett.} {\bf 105} 013903

\bibitem{Buslaev1993}
Buslaev V and Grecchi V 1993 \textit{J. Phys. A} {\bf 26} 5541

\bibitem{Bender2005}
Bender C M 2005 \textit{Contemp. Phys.} {\bf 46} 277

\bibitem{Muga2005}
Ruschhaupt A, Delgado F and Muga J G 2005 \textit{J. Phys. A} {\bf 38} L171

\bibitem{Ganainy2007}
El-Ganainy R, Makris K G, Christodoulides D N and Musslimani Z H 2007 \textit{Opt. Lett.} {\bf 32} 2632

\bibitem{Longhi2009}
Longhi S 2009 \textit{Phys. Rev. Lett.} {\bf 103} 123601

\bibitem{Guo2009}
Guo A, Salamo G J, Duchesne D, Morandotti R, Volatier-Ravat M, Aimez V, Siviloglou G A and Christodoulides D N 2009 \textit{Phys. Rev. Lett.} {\bf  103} 093902

\bibitem{Feng2011}
Feng L, Ayache M, Huang J, Xu Y -L, Lu M -H, Chen Y -F, Fainman Y and Scherer A 2011 \textit{Science} {\bf 333} 729


\bibitem{Alexeeva2012}
Alexeeva N V, Barashenkov I V, Sukhorukov A A and Kivshar Y S 2012 \textit{Phys. Rev. A} {\bf 85} 063837

\bibitem{Regensburger2012}
Regensburger A, Bersch C, Miri M -A, Onishchukov G, Christodoulides D N and Peschel U 2012 \textit{Nature} {\bf 488} 167



\bibitem{Huse2015}
Nandkishore R and Huse D A 2015 \textit{Annu. Rev. Condens. Matter Phys.} {\bf 6} 15

\bibitem{Ponte2015}
Ponte P, Papi\'c Z, Huveneers F and Abanin D A 2015 \textit{Phys. Rev. Lett.} {\bf 114} 140401

\bibitem{Lazarides2015}
Lazarides A, Das A and Moessner R 2015 \textit{Phys. Rev. Lett.} {\bf 115} 030402

\bibitem{Zhang2017}
Zhang J, Hess P, Kyprianidis A, Becker P, Lee A, Smith J, Pagano G, Potirniche I -D, Potter A C and Vishwanath A 2017 \textit{Nature} {\bf 543} 217

\bibitem{Yao2017}
Yao N Y, Potter A C, Potirniche I -D and Vishwanath A 2017 \textit{Phys. Rev. Lett.} {\bf 118} 030401

\bibitem{Yao2018}
Yao N Y, Nayak C, Balents L and Zaletel M P 2020 \textit{Nat. Phys.} {\bf 16} 438

\bibitem{Nayak2018}
Yao N Y and Nayak C 2018 \textit{Phys. Today} {\bf 71} 40

\bibitem{Lindner2011}
Lindner N H, Refael G and Galitski V 2011 \textit{Nat. Phys.} {\bf 7} 490

\bibitem{Titum2015}
Titum P, Lindner N H, Rechtsman M C and Refael G 2015 \textit{Phys. Rev. Lett.} {\bf 114} 056801
\bibitem{Zhou2016}
Zhou L, Chen C and Gong J 2016 \textit{Phys. Rev. B} {\bf 94} 075443
\bibitem{Roy2017}
Roy R and Harper F 2017 \textit{Phys. Rev. B} {\bf 96} 155118 2017


\bibitem{Grifoni1998}
Grifoni M and Hanggi P 1998 \textit{Phys. Rep.} {\bf 304} 229

\bibitem{Della2007}
Valle G Della, Ornigotti M, Cianci E, Foglietti V, Laporta P and Longhi S 2007 \textit{Phys. Rev. Lett.} {\bf 98} 263601

\bibitem{Della2013}
Valle G Della and Longhi S 2013 \textit{Phys. Rev. A} {\bf  87} 022119

\bibitem{Huang2020a}
Huang L  and  Lai Y C  2020 {\it Commun. Theor. Phys.} {\bf 72} 047601

\bibitem{Huang2020b}
Li X L, Chen X Z,  Liu C R and  Huang L  2020 {\it Acta Phys. Sin.} {\bf 69} 080506

\bibitem{Wu2012}
Wu J and Xie X T 2012 \textit{Phys. Rev. A} {\bf 86} 032112
\bibitem{Ganainy2012}
El-Ganainy R, Makris K G and Christodoulides D N 2012 \textit{Phys. Rev. A} {\bf 86} 033813
\bibitem{Moiseyev22011}
Moiseyev N 2011 \textit{Phys. Rev. A} {\bf 83} 052125
\bibitem{Gong2013}
Gong J and Wang Q -h 2013 \textit{J. Phys. A} {\bf 46} 485302

\bibitem{Zhou2018}
Zhou L and Gong J 2018 \textit{Phys. Rev. B} {\bf 98} 205417


\bibitem{West2010}
West C T, Kottos T and Prosen T 2010 \textit{Phys. Rev. Lett.} {\bf 104} 054102

\bibitem{Longhi2017}
Longhi S 2017 \textit{J. Phys. A} {\bf 95} 012125

\bibitem{Zhao19}
Zhao W L, Wang J, Wang X and Tong P 2019 \textit{Phys. Rev. E} {\bf 99} 042201

\bibitem{Haake2010}
Haake F 2010 {\it Quantum Signatures of Chaos} (3rd edn) (Springer-Verlag Berlin Heidelberg) p. 247

\bibitem{Casati1979}
Casati G, Chirikov B V, Izraelev F M and Ford J 1979 {\it Stochastic Behavior in Classical and Quantum Hamiltonian Systems, Lecture Notes
in Physics} (Springer, Berlin) p. 334

\bibitem{Casati1987}
Casati G, Chirikov B V, Shepelyansky D L and Guarneri I 1987 \textit{Phys. Rep.} {\bf 154} 77

\bibitem{Izrailev1990}
Izrailev F M 1990 \textit{Phys. Rep.} {\bf 196} 299

\bibitem{Wang2011}
Wang J, Guarneri I, Casati G and Gong J 2011 \textit{Phys. Rev. Lett.} {\bf 107} 234104

\bibitem{Liujie06}
Liu J, Zhang C, Raizen M G and Niu Q 2006 \textit{Phys. Rev. A} {\bf 73} 013601

\bibitem{Fishman1982}
Fishman S, Grempel D R and Prange R E 1982 \textit{Phys. Rev. Lett.} {\bf 49} 509

\bibitem{Podolskiy2004}
Podolskiy V A, Narimanov E, Fang W and Cao H 2004 \textit{Proc. Natl. Acad. Sci. USA} {\bf 101} 10498

\bibitem{Raizen1999}
Raizen M G and Steck D L 2011 \textit{Scholarpedia} {\bf 6} 10468 (2011)

\bibitem{Rosen2000}
Rosen A, Fischer B, Bekker A and Fishman S 2000 \textit{J. Opt. Soc. Am. B} {\bf 17} 1579


\bibitem{Chaudhury2009}
Chaudhury S, Smith A, Anderson B, Ghose S and Jessen P 2009 \textit{Nature} {\bf 461} 768

\bibitem{Zhao20}
Zhao W L and Jie Q L 2020 \textit{Chin. Phys. B} {\bf 29} 080302

\bibitem{Zhao2010}
Zhao W L, Jie Q L and Zhou B 2010 \textit{Commun. Theor. Phys.} {\bf 54} 247

\bibitem{Zhao09}
Zhao W L and Jie Q L 2009 \textit{Commun. Theor. Phys.} {\bf 51} 465


\bibitem{Yang15}
Yang Y B and Wang W G 2015 \textit{Chin. Phys. Lett.} {\bf 32} 030301


\bibitem{Graefe15}
Graefe E -M, Korsch H J, Rush A and Schubert R 2015 \textit{J. Phys. A} {\bf 48} 055301

\bibitem{Bender2009}
Bender C M, Feinberg J, Hook D W and Weir D J  2009 \textit{Pramana J. Phys.} {\bf 73} 453

\bibitem{Bender09bk}
Bender C M and Brody D C 2009
\textit{Optimal Time Evolution for Hermitian and Non-Hermitian Hamiltonians, Lecture Notes in Physics} (Springer, Berlin Heidelberg) \textbf{789} 341

\bibitem{Bender99}
Bender C M, Boettcher S and Meisinger P N 1999 \textit{J. Math. Phys.} {\bf 40} 2201


\bibitem{Bender11}
Bender C M and Hook D W 2011 \textit{J. Phys. A} {\bf 44} 372001


\bibitem{Larkin1969}
Larkin A and Ovchinnikov Y N 1969 \textit{Sov. Phys. JETP } {\bf 28} 1200

\bibitem{Maldacena2016}
Maldacena J, Shenker S H and Stanford D 2016 \textit{J. High Energ. Phys.} {\bf 2016} 106


\bibitem{Hashimoto2017}
Hashimoto K, Murata K and Yoshii R 2017 \textit{J. High Energ. Phys.} {\bf 2017} 138

\bibitem{Dora2017}
D\'ora B and Moessner R 2017 \textit{Phys. Rev. Lett.} {\bf 119} 026802


\bibitem{Heyl2018}
Heyl M, Pollmann F and D\'ora B 2018 \textit{Phys. Rev. Lett.} {\bf 121} 016801

\bibitem{Mata2018}
Garc\'ia-Mata I, Saraceno M, Jalabert R A, Roncaglia A J and Wisniacki D A 2018 \textit{Phys. Rev. Lett.} {\bf 121} 210601

\bibitem{Mata2018B}
Jalabert R A, Garc\'ia-Mata I and Wisniacki D A 2018 \textit{Phys. Rev. E} {\bf 98} 062218

\bibitem{Fortes2019}
Fortes E M, Garc\'ia-Mata I, Jalabert R A and Wisniacki D A 2019 \textit{Phys. Rev. E} {\bf 100} 042201

\bibitem{Ueda2018}
Hamazaki R, Fujimoto K and Ueda M \textit{Operator Noncommutativity and Irreversibility in Quantum Chaos} arXiv:1807.02360 [condmat. stat-mech]

\bibitem{Yan19}
Yan H, Wang J Z and Wang W G 2019 \textit{Commun. Theor. Phys.} {\bf 71} 1359

\bibitem{Swingle2016}
Swingle B, Bentsen G, Schleier-Smith M and Hayden P 2016 \textit{Phys. Rev. A} {\bf 94} 040302(R)

\bibitem{Hafezi2016}
Zhu G, Hafezi M and Grover T 2016 \textit{Phys. Rev. A} {\bf 94} 062329

\bibitem{Li2017}
Li J, Fan R, Wang H, Ye B, Zeng B, Zhai H, Peng X and Du J 2017 \textit{Phys. Rev. X} {\bf 7} 031011

\bibitem{Garttner2017}
G\"arttner M, Bohnet J G, Safavi-Naini A, Wall M L, Bollinger J J and Rey A M 2017 \textit{Nat. Phys.} {\bf 13} 781

\bibitem{Chirikov1979}
Chirikov B V 1979 \textit{Phys. Rep.} {\bf 52} 263

\bibitem{analysis}
See Appendix for detailed derivations  of the expomentially-fast growth of a specific trajectory. There is the derivaiton of the tangent mapping equaitons of the deviation of two trajectories.


\bibitem{Rozenbaum17}
Rozenbaum E B, Ganeshan S and Galitski V 2017 \textit{Phys. Rev. Lett.} {\bf 118} 086801




\end{thebibliography}
\end{document}